# Killig-Yano Tensors of Valence-2 in the Standart Kaluza-Klein Theory

Ali Nur Nurbaki[1]

*Department of Physics, Middle East Technical University, Ankara, Turkey*

This work is constructed on two main concepts: Killing-Yano symmetry and the Kaluza-Klein theory. Those concepts are reviewed in the first three chapters. In the fourth chapter firstly Killing-Yano equations of valence-2 are obtained for a general 5-dimensional metric and then reduced for the Minkowski metric. It is seen that in order to save the existence of Killing-Yano tensors with the fifth components some conditions must be satisfied.

---

[1] 136379@metu.edu.tr

# Contents

# CHAPTER 1

# INTRODUCTION

## BACKGROUND, MOTIVATION AND OUTLINE

Unification of fundamental forces has always been a unique and idealistic problem in physics. In the past hundred years, there have been many attempts to join the fundamental forces in one theory. Today most of these attempts turned out to be of little or no use.

After the discovery of general relativity, attempts to unify gravity with other fundamental forces were boosted. Although it was not considered to be successful the first courageous trial was made by Herman Weyl in 1918. Later on, in 1926 Kaluza suggested a 5-D metric with off diagonal components of the fifth dimension to be the vector potential of the electromagnetic theory. Writing Einstein's equations in this structure made the theory split into two fundamental theories: the 4-D theory of relativity and again Maxwell's electromagnetic theory in four dimensions. In 1926, Klein elaborated on the structure of the extra dimension and suggested that the fourth spatial dimension is curled up in a circle of a very small radius. Kaluza –Klein theory was an original and a revolutionary idea. However, because of the pure geometric understanding of the theory of relativity and so that of the Kaluza-Klein theory an incompatibility with the quantum theory has aroused. This was due to the fact that quantum theory was perceiving forces as fields, not pure geometry. Because of this, the quantization of gravity did not conform to classical relativity.

This compatibility problem with the quantum field theory and the new understanding of nuclear forces made Kaluza Klein theory forgotten. On the other hand being the mother of all extra dimensional theories, it has always received an important place among all unification theories.

Supergravity and superstring theories, which as of today are known to be the best surviving models for quantum gravity, unify all fundamental forces. Supersymmetric theories are known to be invariant under some transformations called supersymmetry transformations. Such theories admit some superinvariants called supercharges. The supercharges that are admitted not by the structure of the model itself but by the background metric are called non-generic supercharges. It is known that such supercharges are related to some hidden symmetries of space-time called Killing-Yano tensors [1]. If the metric admits such tensors then it can be inferred that there exist supercharges related to those tensors.

A Killing-Yano tensor can also be defined as the square-root of a Killing tensor, in the sense that a second rank Killing tensor can be constructed by appropriately contracting two KY tensors of rank two or more.

In this thesis we are going to investigate Killing-Yano (KY) tensors of valence two in the 5-D Kaluza-Klein (KK) theory.

The second chapter is about space-time symmetries and isometries. Killing vectors, Killing tensors and Killing-Yano tensors will be introduced. The relations of these concepts among each other will be clarified. The metric for plane fronted waves with parallel rays will be introduced. KY tensors of rank two and three will be investigated for this background metric.

In the third chapter we are going to introduce Kaluza-Klein theory briefly in an anholonomic basis called the "horizontal lift basis" which makes the metric diagonal in its fifth components. For this purpose we shall give the basics of Einstein's equations in 4-D.

In the fourth chapter we are going to write KY equations in five dimensions. Then using the KK reduction mechanism the expressions of the KY equations in four dimensions will be given. As a special case, the KY equations, where the 4-D metric is Minkowskian, will be examined.

The final chapter will be devoted to discussions and conclusions.

## NOTATIONS AND CONVENTIONS

The notation we are going to use is as follows. All capital Latin letters A,B,C are used for five dimensional indices, 0,1,2,3,5, and normal Greek letters μ,λ,ν are used for the common 4–D indices, 0,1,2,3. All quantities with a hat are again in five–dimensions. We used natural units for constants, and thus they are all equal to one. The metric signature is (-,+,+,+,+) in five dimensions.

In the following chapters we shall use the notation

$K_{(ab,c)}$ ,

where the semicolon denotes the covariant derivative and parentheses denote symmetrization.

We shall also use the notation $D_\alpha X_{\beta\lambda}$ for covariant derivative.

# CHAPTER 2

# ISOMETRIES, CONSERVATION LAWS, DUAL METRICS AND HIDDEN SYMMETRIES OF SPACE-TIME

## ISOMETRIES AND KILLING VECTORS

Lorenz invariance which is the main symmetry related to special theory of relativity, turns out to be local in general relativity. Therefore, symmetries that are not local are valuable in general relativity. Isometries are good candidates for such types of symmetries if they exist. An isometry is a transformation that leaves the metric invariant. They are generated by vector fields called Killing vector fields such that they satisfy

$$D_\nu X_\mu + D_\mu X_\nu = 0 \qquad (2.1.1)$$

or in other words

$$L_X g_{\mu\nu} = 0 \qquad (2.1.2)$$

meaning that the metric is form invariant when dragged along a Killing vector field. This means that length between points and angles between tangent vectors on a curve are preserved when dragged along a Killing vector field.

In the light of these facts, it is possible to observe a close relation between conservation laws and Killing vectors.

When the components $g_{\mu\nu}$ relative to some coordinate basis are independent of some coordinate, then the tangent vector corresponding to that coordinate is a Killing vector.

We can understand the corresponding conservation laws with the help of Hamiltonian mechanics. Killing vectors correspond to cyclic coordinates in Hamilton mechanics. So that the momentum corresponding to that cyclic coordinate is conserved. Thus, we can conclude that motion along a Killing vector field leaves the momentum conserved.

## KILLING TENSORS AND DUAL METRICS

It is possible to make some generalizations on conserved quantities, isometries and Killing vectors. Killing vectors have been generalized to second rank Killing tensors such that they satisfy
$$K_{(\mu\nu,\lambda)} = 0. \qquad (2.2.1)$$

Construction of Killing tensors produce some symmetries, where they can be used to simplify equations of general relativity [21]. They also play an important role in the complete solution of the geodesic equation. Killing tensor equation simplifies in some particular space-times, which is another interesting feature of Killing tensors [22],[23]. Recently, new methods are developed to find Killing tensors for some space-times possessing other types of symmetries [24]. The motion of a particle moving on a geodesic in a space admitting isometries is of great importance in understanding the structure of its underlying manifold.

The Lagrangian for such a particle can be written in the form

$$L = (1/2) g_{\mu\nu} \dot{x}^\mu \dot{x}^\nu. \tag{2.2.2}$$

Using this Lagrangian, Hamiltonian can be written with the help of the contravariant metric,

$$H = (1/2) g^{\mu\nu} P_\mu P_\nu. \tag{2.2.3}$$

The time evolution of any function on the phase space can be calculated via the Poisson bracket of that function with the Hamiltonian,

$$(dK/d\tau) = \{K, H\}. \tag{2.2.4}$$

Here basic Poisson brackets are

$$\{x^\mu, p_\nu\} = \delta_\nu^\mu. \tag{2.2.5}$$

The contravariant Killing tensor can be defined as

$$K^{\mu\nu} = g^{\mu\lambda} K_{\lambda\alpha} g^{\alpha\nu}. \tag{2.2.6}$$

This can be used to construct a conserved quantity quadratic in momenta such that

$$K = (1/2) K^{\mu\nu} P_\mu P_\nu. \tag{2.2.7}$$

In view of the equations (2.2.5) and (2.2.7), the Poisson bracket of K with the Hamiltonian vanishes

$$\{K, H\} = 0. \tag{2.2.8}$$

So this makes K a constant of motion, which generates symmetry transformations linear in momentum

$$\{x^\mu, K\} = K^{\mu\nu} P_\nu. \tag{2.2.9}$$

Since H and K are both quadratic in momenta, it can be easily seen that K and H are formally similar. Motivated by this formal similarity, a duality relation between $g^{\mu\nu}$ and $K^{\mu\nu}$ can be sought.

Two models can be constructed from two points of view [2]:

In the first model the Hamiltonian is H and constant of motion is K as usual, whereas in the second model the Hamiltonian is K and the constant of motion is H.

These two models give rise to the following interesting results.

If contravariant Killing tensors are generated by the inverse metric, then $K^{\mu\nu}$ can be thought as the contravariant components of a metric which generate Killing tensor $g^{\mu\nu}$.

Then

$$K^{\mu\nu} \equiv \tilde{g}^{\mu\nu} \tag{2.2.10}$$

such that

$$\tilde{g}^{\mu\nu} \tilde{g}_{\lambda\nu} = \delta^\mu_{\ \nu}. \tag{2.2.11}$$

So, $\tilde{g}_{\mu\nu}$ is the covariant metric and consequently it must be covariantly constant,

$$\tilde{D}_\lambda \tilde{g}_{\mu\nu} = 0. \tag{2.2.12}$$

Using the connection $\tilde{\Gamma}^\lambda_{\mu\nu}$ and

$$\tilde{K}_{\mu\nu} \equiv g_{\mu\nu}, \tag{2.2.13}$$

will be the Killing tensor with respect to the metric $\tilde{g}_{\mu\upsilon}$ which satisfies

$$\tilde{D}_{(\lambda}\tilde{K}_{\mu\nu)} = 0. \qquad (2.2.14)$$

This duality should be interpreted as follows: Second rank Killing tensors can be thought of as metrics in another space dual to the original metric space such that in that space, original metric becomes a second rank Killing tensor. Mapping a Killing tensor to a metric by this way twice, brings the original metric back [2].

## HIDDEN SYMMETRIES AND KILLING-YANO TENSORS

After some study of isometries, Killing vectors and their generalizations which are Killing tensors, some hidden symmetries can be understood better.

A good candidate for a hidden symmetry is the well known Killing-Yano symmetry.
A Killing-Yano tensor of valence *n* is an anti-symmetric tensor satisfying

$$f_{\upsilon_1\upsilon_2(\upsilon_n,\lambda)} = 0. \qquad (2.3.1)$$

The relation between Killing tensors and Killing-Yano tensors is that Killing-Yano tensors can be considered as the square-root of Killing tensors in the following way,

$$\mathrm{K}^{\mu\upsilon} = g^{\alpha\beta} f_{\mu\alpha} f_{\beta\nu}, \qquad (2.3.2)$$

for second rank KY tensors, and for third rank KY tensors we have

$$K_{\mu\nu} = g^{\alpha\delta} g^{\beta\gamma} f_{\mu\alpha\beta} f_{\gamma\delta\nu}. \qquad (2.3.3)$$

Killing-Yano tensors are important for field theories. For example existence of Killing-Yano tensors are closely related to the separability of Dirac equation in Kerr space-times [3].

Another feature of Killing-Yano tensors is that Killing-Yano tensors of valence three are known to be related with Lax tensors [4].

But the most important feature of Killing-Yano tensors is that they are important for spinning particle models on curved backgrounds, which is a supersymmetric model [5].

Supersymmetry is still a popular modern idea for theoretical physics. A supersymmetric theory means that the theory is invariant under supersymmetry transformations. Supersymmeric models are used to admit some invariants called supercharges.

Killing-Yano tensor of valence $n$ admits some kind of supersymmetries called non-generic supercharges. Non-generic supercharges and the generic supercharges are different from each other, in the sence that the generic supercharges are about the theory itself while the non-generic ones are metric dependent. Killing-Yano tensors are metric dependent and they admit non-generic supercharges in the following way.

A supercharge related to a Killing-Yano tensor of valence $n$ is defines as:

$$Q_f = f_{\nu_1 \nu_2 ... \nu_n} \Pi^{\nu_1} \psi^{\nu_2} ... \psi^{\nu_n} . \qquad (2.3.4)$$

This is a superinvariant because of the bracket

$$\{Q_0, Q_f\} = 0 . \qquad (2.3.5)$$

Here $Q_0 = \Pi_\mu \psi^\mu$, $\Pi_\mu$ is called covariant momenta and $\psi^\mu$ are called odd Grassman variables. These are anti-commuting numbers with each other, but commuting with ordinary numbers, such that

$$\psi^\mu \psi^\nu = -\psi^\nu \psi^\mu \qquad (2.3.6)$$

$$\psi^\mu x = x\psi^\mu. \tag{2.3.7}$$

By using the Hamiltonian

$$H = (1/2)g^{\mu\upsilon}\Pi_\mu \Pi_\upsilon, \tag{2.3.8}$$

Jacobi identities and the Killing-Yano equation, one can see that this supercharge is also a constant of motion too, i.e.

$$\left[Q_f, H\right] = 0. \tag{2.3.9}$$

Now, in the next section, an example about finding Killing-Yano tensors of valence-2 for the 4-D pp-wave metric is going to be examined.

In the fourth chapter Killing-Yano tensors of valence two are going to be investigated. But this case will have some differences from the standard approach in the sense that Killing-Yano equations are to be obtained within the framework of the Kaluza-Klein theory.

## KILLING-YANO TENSORS FOR PP-WAVE METRIC

After introducing the importance of Killing and Killing-Yano tensors we make a calculation for a particular class of metric i.e. the pp-wave metric. Such a calculation was made in 2002 [6], [12].

The pp-wave metric is one of the important classes of metrics which satisfy Einstein's equations. Its importance comes from the fact that this metric admit plane fronted waves with parallel rays. The propagated radiation can be either electromagnetic or gravitational.

The famous pp-wave in four dimensions can be introduced in the form [7]

$$ds^2 = 2dvdu + dx^2 + dy^2 + H(x,y,u)du^2. \tag{2.4.1}$$

Here "u" and "v" are coordinates defined in terms of both space and time variables, but "v" does not seem to exhibit a physical meaning; it stands for an affine parameter, while "u" is called the retarded time.

For this metric KY equations are explicitly found to be as:

$$f_{12,v} = 0 \tag{2.4.2}$$

$$f_{13,v} = 0 \tag{2.4.3}$$

$$f_{14,v} = 0 \tag{2.4.4}$$

$$f_{12,x} = 0 \tag{2.4.5}$$

$$f_{13,y} = 0 \tag{2.4.6}$$

$$f_{23,y} = 0 \tag{2.4.7}$$

$$f_{23,v} + f_{13,x} = 0 \tag{2.4.8}$$

$$f_{24,v} + f_{14,x} = 0 \tag{2.4.9}$$

$$f_{23,v} - f_{12,y} = 0 \tag{2.4.10}$$

$$f_{34,v} + f_{14,y} = 0 \tag{2.4.11}$$

$$f_{24,v} - f_{12,u} = 0 \tag{2.4.12}$$

$$f_{34,v} - f_{13,u} = 0 \tag{2.4.13}$$

$$f_{13,x} + f_{12,y} = 0 \tag{2.4.14}$$

$$f_{14,v} + f_{12,u} = 0 \tag{2.4.15}$$

$$f_{14,y} + f_{13,u} = 0 \tag{2.4.16}$$

$$h_{,x} f_{12} + 2 f_{24,x} = 0 \tag{2.4.17}$$

$$h_{,y} f_{13} + 2 f_{34,y} = 0 \tag{2.4.18}$$

$$h_{,x} f_{12} + h_{,y} f_{13} + 2 f_{14,u} = 0 \tag{2.4.19}$$

$$h_{,x} f_{13} + h_{,y} f_{12} + 2 f_{34,x} + 2 f_{24,y} = 0 \tag{2.4.20}$$

$$h_{,u} f_{12} - h_{,x} f_{14} + h_{,y} f_{23} + 2 f_{24,u} = 0 \tag{2.4.21}$$

$$h_{,u} f_{13} - h_{,y} f_{14} - h_{,x} f_{23} + 2 f_{34,y} = 0 \tag{2.4.22}$$

$$h_{,x} f_{13} - 2 h_{,y} f_{12} - 2 f_{23,u} - 2 f_{24,y} = 0 \tag{2.4.23}$$

$$h_{,y} f_{12} - 2 h_{,x} f_{13} + 2 f_{23,u} - 2 f_{34,x} = 0 \tag{2.4.24}$$

There is no a priori way to determine whether KY equations admit six non-zero KY tensor components for a given form of a metric. An optimum solution should have a maximum number of non-zero components of the KY tensor, while imposing as less restriction as possible on the metric function. It is found that the KY equations (2.4.2)-(2.4.24) admit a two component non-vanishing Killing-Yano tensor. These are [12]:

$$f_{24} = c_1, \qquad f_{34} = c_2, \qquad (2.4.25)$$

with no restrictions on the metric function $h(x, y, u)$. Using (2.3.2) and (2.4.25) the corresponding Killing tensor is found to have only one non-vanishing component

$$K_{44} = c_1^2 + c_2^2. \qquad (2.4.26)$$

Since there aren't any surviving components, this tensor does not admit a dual metric.

Killing-Yano tensors of valence three are also calculated with maximum number of components i.e. three, but with some restrictions on the metric function $h(x, y, u)$. If the function is to satisfy the equation below

$$2\partial_u q(u) - c_2 \partial_x h(x, y, u) + c_1 \partial_y h(x, y, u) = 0, \qquad (2.4.27)$$

where $q(u)$ is an arbitrary function, then the non-vanishing components are found to be

$$f_{134} = c_2, \qquad f_{124} = c_1, \qquad f_{234} = q(u). \qquad (2.4.28)$$

So (2.4.27) has the solution

$$h(x, y, u) = F(c_1 x + c_2 y) + \partial_u q(u)(\frac{y}{c_1} - \frac{x}{c_2}). \qquad (2.4.29)$$

In order to have three component solutions for Killing-Yano tensors of valence three for the pp-wave metric, equation (2.4.29) gives the restriction on the metric.

The Killing tensor components corresponding to (2.4.28) are calculated through the relation (2.3.3), and they are found to be

$$K_{14} = -2(c_1^2 + c_2^2), \qquad K_{24} = -2c_2 q(u),$$

$$K_{22} = -2c_1^2, \qquad K_{23} = -2c_1 c_2, \qquad (2.4.30)$$

$$K_{33} = 2c_2^2, \qquad K_{34} = -2c_1 q(u),$$

$$K_{44} = 2(q(u)^2 - (c_1^2 + c_2^2)h(x, y, u)).$$

Again this Killing tensor doesn't admit a dual metric because the contravariant components as defined in (2.2.6) turn out to be a singular matrix.

So the admission of dual metrics is a special situation for Killing symmetries.

# CHAPTER 3

# 5-D KALUZA-KLEIN THEORY WITH A

# SIMPLER BASE

Making physics with more than 4-dimensions is not that much of a new idea indeed. Nearly a hundred years ago in 1914 Nordström realized that electromagnetism and gravity can be combined in 5-dimensions [19], but later on this theory of gravity was forgotten. In 1921 when Theodor Kaluza published his paper [14] "On the unification problem in physics", the idea of the fifth dimension was revoked.

His main assumptions were:

1) To extend General relativity to 5-dimensions.
2) Fifth coordinate shouldn't be observed.

Later on O. Klein suggested that fifth dimension was circular [17], [18], then the second assumption was verified. Extra dimensional theories are one of the most important instruments for theoretical physics. Today many physicist concentrate on the relativistic structure of Kaluza-Klein theories as well as its particle physics aspect [20]. But in order to introduce Kaluza-Klein theory, we should first go over the conventional theory of General Relativity and its Riemannian structure briefly.

**PRELIMINARIES: GENERAL RELATIVITY IN 4-D AND NOTATIONS**

It is not an overstatement that General Relativity means Einstein's equations. But before introducing them let us define the Riemann curvature tensor and its symmetry properties. Riemann tensor is defined as follows

$$R^{\alpha}{}_{\mu\beta\nu} = \partial_{\beta}\Gamma^{\alpha}{}_{\mu\nu} - \partial_{\nu}\Gamma^{\alpha}{}_{\mu\beta} + \Gamma^{\gamma}{}_{\mu\nu}\Gamma^{\alpha}{}_{\gamma\beta} - \Gamma^{\gamma}{}_{\mu\beta}\Gamma^{\alpha}{}_{\gamma\nu}, \qquad (3.0.1)$$

where

$$\Gamma^{\alpha}{}_{\beta\gamma} = \frac{1}{2}g^{\alpha\delta}(\partial_{\beta}g_{\delta\gamma} + \partial_{\gamma}g_{\beta\delta} - \partial_{\delta}g_{\beta\gamma}), \qquad (3.0.2)$$

are the well known metric connections. Riemann tensor is the curvature tensor in curved geometries and it has some useful symmetries. First of all it is anti-symmetric on its last two indices i.e.,

$$R^{\alpha}{}_{\mu\beta\nu} = -R^{\alpha}{}_{\mu\nu\beta}. \qquad (3.0.3)$$

Odd permutations of its last three index has the following cyclic property

$$R^{\alpha}{}_{\mu\beta\nu} + R^{\alpha}{}_{\nu\mu\beta} + R^{\alpha}{}_{\beta\nu\mu} \equiv 0. \qquad (3.0.4)$$

This identity arises from the fact that connection is symmetric. By using the geodesic coordinates and lowering the first index via the metric

$$g_{\delta\alpha}R^{\alpha}{}_{\mu\beta\nu} = R_{\delta\mu\beta\nu}, \qquad (3.0.5)$$

it can be seen that the Riemann tensor is symmetric under the operation of interchanging the first two with the last two indices

$$R_{\alpha\mu\beta\nu} = R_{\beta\nu\alpha\mu} \;.$$

Using (3.0.5) together with (3.0.3) gives us the following result

$$R_{\alpha\mu\beta\nu} = -R_{\nu\beta\alpha\mu} \;. \tag{3.0.6}$$

Working out all these symmetries for the covariant Riemann tensor we get the equivalent form of the identity (3.0.4)

$$R_{\alpha\beta\gamma\delta} + R_{\alpha\delta\beta\gamma} + R_{\alpha\gamma\delta\beta} \equiv 0. \tag{3.0.7}$$

When the covariant derivative of the Riemann tensor is taken into account, it again satisfies cyclic permutation which looks like (3.0.7)

$$D_\alpha R_{\delta\varepsilon\beta\gamma} + D_\gamma R_{\delta\varepsilon\alpha\beta} + D_\beta R_{\delta\varepsilon\gamma\alpha} \equiv 0, \tag{3.0.8}$$

and are called the Bianchi identities.
Providing sufficient amount about the properties of the Riemann tensor, it is now possible to define the famous Ricci tensor and the Ricci scalar.

Ricci tensor can be defined in terms of Riemann tensor with the following contraction

$$g^{\beta\alpha} R_{\alpha\mu\beta\nu} = R_{\mu\nu} \tag{3.0.9}$$

which is equivalent to

$$R_{\mu\nu} = R^\alpha{}_{\mu\alpha\nu} \tag{3.0.10}$$

or in terms of connections it can be expressed as,

$$R_{\mu\nu} = \partial_\alpha \Gamma^\alpha{}_{\nu\mu} - \partial_\nu \Gamma^\alpha{}_{\alpha\mu} + \Gamma^\alpha{}_{\alpha\delta}\Gamma^\delta{}_{\nu\mu} - \Gamma^\alpha{}_{\nu\delta}\Gamma^\delta{}_{\alpha\mu}. \qquad (3.0.11)$$

The Ricci scalar is defined through the contraction

$$R = g^{\mu\nu} R_{\mu\nu}. \qquad (3.0.12)$$

So we are now ready to define the Einstein's tensor by using $R$ and $R_{\mu\nu}$

$$G_{\mu\nu} = R_{\mu\nu} - \frac{1}{2} R g_{\mu\nu}, \qquad (3.0.13)$$

which can be derived from the Einstein-Hilbert action

$$S = \frac{1}{16\pi G} \int_M d^4 x (\sqrt{-g}) R, \qquad (3.0.14)$$

by taking the variation with respect to the metric $g_{\mu\nu}$. Here $g$ is the determinant of the metric. The covariant derivative of (3.0.13) also satisfy the contracted Bianchi identities [13]

$$D_\mu G^\mu{}_\nu \equiv 0. \qquad (3.0.15)$$

We can define the space-time curvature by matter fields via the well known Einstein's equations:

$$G_{\mu\nu} = 8\pi T_{\mu\nu}, \qquad (3.0.16)$$

where $T_{\mu\nu}$ is the energy momentum tensor which is used to define the distribution of the matter fields through the space-time. Equation (3.0.15) states that

$$D_\mu T^\mu{}_\nu \equiv 0, \qquad (3.0.17)$$

is a manifestation of conservation of energy and momentum.

# THE FIVE-DIMENSIONAL RELATIVITY

Recalling Kaluza's main assumptions, one can extend General Relativity to 5-dimensions by deriving Einstein's equations in 5-dimensions. For this purpose we can use a 5-D Einstein-Hilbert action such that

$$S = \frac{1}{16\pi \hat{G}} \int_M d^4x\, dy (\sqrt{-\hat{g}})\hat{R}. \tag{3.1.1}$$

Here y stands for the fifth coordinate $x^5$. Varying this action with respect to the 5-D metric $\hat{g}_{AB}$, we derive 5-D Einstein equations for vacuum as

$$\hat{G}_{AB} = 0, \tag{3.1.2}$$

where

$$\hat{G}_{AB} = \hat{R}_{AB} - \frac{1}{2}\hat{R}\hat{g}_{AB}. \tag{3.1.3}$$

As expected, here $\hat{R}_{AB}$ is 5-D Ricci tensor, $\hat{R}$ is 5-D Ricci scalar and $\hat{g}_{ab}$ is the 5-D Kaluza metric. Using (3.1.2) and (3.1.3) and contracting by $\hat{g}^{AB}$, the contravariant metric, we get

$$\hat{R}_{AB} = 0. \tag{3.1.4}$$

This is called the Ricci flat condition and the solutions of (3.1.4) are vacuum metrics.

But we should be interested in the full 5-D equations with energy stress tensor at the right hand side:

$$\hat{G}_{AB} \equiv 8\pi \hat{G} \hat{T}_{AB}. \tag{3.1.5}$$

In the next sections we are going to get these equations in terms of both Einstein's equations and Maxwell stress tensor.

## ELECTROMAGNETIC STRUCTURE OF THE 5-D METRIC

The 5-D metric tensor defined above can be written in the form

$$\hat{g}_{AB} = \begin{pmatrix} \hat{g}_{\mu\nu} & \hat{g}_{\mu 5} \\ \hat{g}_{5\nu} & \hat{g}_{55} \end{pmatrix}. \tag{3.2.1}$$

Due to the coordinate-dependent structure of the metric tensor, Kaluza expressed the $\hat{g}_{\mu 5}$ components in terms of the 4-potentials $A_\mu$, and define $\hat{g}_{55}$ by a scalar field $\varphi$, so that he could build a metric tensor including both gravitation and electromagnetism.

With a suitable choice of coordinates, such a 5-D metric can be constructed as follows [14], [16], [17]

$$\hat{g}_{AB} = \begin{pmatrix} g_{\mu\nu} + \gamma^2 \varphi^2 A_\mu A_\nu & \gamma \varphi^2 A_\mu \\ \gamma \varphi^2 A_\nu & \varphi^2 \end{pmatrix}, \tag{3.2.2}$$

which has the contravariant form

$$\hat{g}^{AB} = \begin{pmatrix} g^{\mu\nu} & -\gamma A^\mu \\ -\gamma A^\nu & \varphi^{-2} + \gamma^2 A_\mu A^\mu \end{pmatrix}. \tag{3.2.3}$$

Now the well known line element has the infinitesimal form

$$ds^2 = \hat{g}_{\mu\nu}dx^\mu dx^\nu + \hat{g}_{\mu 5}dx^\mu dy + \hat{g}_{5\nu}dydx^\nu + \hat{g}_{55}dydy, \qquad (3.2.4)$$

and can be written in terms of the metric components as

$$ds^2 = g_{\mu\nu}dx^\mu dx^\nu + \varphi^2(dy + \gamma A_\mu dx^\mu)^2, \qquad (3.2.5)$$

where $A_\mu$ is a 4-vector of the form

$$A_\mu = (\phi, \vec{A}). \qquad (3.2.6)$$

Here $\gamma$ is a scaling factor and can be chosen to be "1".

### REDUCED FIELD EQUATIONS

Since the Ricci tensor $\hat{R}_{AB}$ is symmetric then for 5-D it has 15 independent components, meaning that

$$\hat{R}_{AB} = 0, \qquad (3.3.1)$$

yields 15 field equations. Using the definition of the Ricci tensor, 5-D Ricci tensor can be constructed as

$$\hat{R}_{AB} = \partial_C \hat{\Gamma}^C_{AB} + \partial_B \hat{\Gamma}^C_{AC} + \hat{\Gamma}^C_{AB}\hat{\Gamma}^D_{CD} - \hat{\Gamma}^C_{AD}\hat{\Gamma}^D_{BC}, \qquad (3.3.2)$$

where $\hat{\Gamma}^C_{AB}$ `s are the well known three-index symbols which are derived from the metric via

$$\hat{\Gamma}^C_{AB} = \frac{1}{2}\hat{g}^{CD}(\partial_A \hat{g}_{DB} + \partial_B \hat{g}_{DA} - \partial_D \hat{g}_{AB}). \qquad (3.3.3)$$

The key instrument that Kaluza used to construct the theory was the structure of these three-index symbols with $5^{th}$ components. These can be calculated by using (3.2.2), (3.2.3), (3.3.3) and the cylinder condition which assumes that all derivatives with respect to $5^{th}$ dimension should vanish.

This fact makes $\partial_y$ a Killing vector field.

So we get the connections with fifth components as follows

$$\hat{\Gamma}^{\mu}_{\nu\lambda} = \Gamma^{\mu}_{\nu\lambda} - \frac{1}{2}\varphi^2(A_\nu F^{\mu}_{\ \lambda} + A_\lambda F^{\mu}_{\ \nu}) - \varphi\partial^{\mu}\varphi A_\nu A_\lambda,$$

$$\hat{\Gamma}^{\mu}_{\nu 5} = -\frac{1}{2}\varphi^2 F^{\mu}_{\ \nu}\varphi\partial^{\mu}\varphi A_\nu$$

$$\hat{\Gamma}^{5}_{\mu\nu} = \frac{1}{2}(D_\mu A_\nu + D_\nu A_\mu) - \frac{1}{2}\varphi^2 A^{\lambda}(A_\mu F_{\nu\lambda} + A_\nu F_{\mu\lambda}) +$$

$$A_\mu A_\nu \varphi \partial^{\lambda}\varphi A_\lambda + \frac{1}{\varphi}(A_\mu \partial_\nu \varphi + A_\nu \partial_\mu \varphi),$$
(3.3.4)

$$\hat{\Gamma}^{\mu}_{55} = -\varphi\partial^{\mu}\varphi,$$

$$\Gamma^{5}_{5\nu} = \frac{1}{2}\varphi^2 A^{\lambda} F_{\lambda\nu} + \varphi\partial^{\lambda}\varphi A_\lambda A_\nu + \frac{1}{\varphi}\partial_\nu\varphi,$$

$$\Gamma^{5}_{55} = \varphi\partial^{\lambda}\varphi A_\lambda.$$

Also

$$\hat{\Gamma}^{A}_{\mu A} = \hat{\Gamma}^{\lambda}_{\mu\lambda} + \hat{\Gamma}^{5}_{\mu 5} = \Gamma^{\lambda}_{\mu\lambda} + \frac{1}{\varphi}\partial_\mu\varphi,$$
(3.3.5)
$$\hat{\Gamma}^{A}_{5A} = \hat{\Gamma}^{\lambda}_{5\lambda} + \hat{\Gamma}^{5}_{55} = 0,$$

where

$$F_{\mu\nu} = \partial_\mu A_\nu - \partial_\nu A_\mu$$
(3.3.6)

is the Maxwell tensor.
This tensor can be written in the following matrix form

$$F_{\mu\nu} = \begin{pmatrix} 0 & E_x & E_y & E_z \\ -E_x & 0 & -B_z & B_y \\ -E_y & B_z & 0 & -B_x \\ -E_z & -B_y & B_x & 0 \end{pmatrix}. \tag{3.3.7}$$

By using (3.3.2), (3.3.3), (3.3.4), (3.3.5) we can obtain the reduced form of $\hat{R}_{AB}$ as

$$\begin{aligned}
\hat{R}_{\mu\nu} &= R_{\mu\nu} - \frac{1}{2}\varphi^2 F_{\mu\lambda}F_\nu{}^\lambda - \frac{1}{\varphi}(D_\nu\partial_\mu\varphi) - \frac{1}{2}A_\mu[\varphi^2(D_\lambda F^\lambda{}_\nu) + 3\varphi\partial^\lambda\varphi F_{\lambda\nu}] \\
&\quad -\frac{1}{2}A_\nu[\varphi^2(D_\lambda F^\lambda{}_\nu) + 3\varphi\partial^\lambda\varphi F_{\lambda\mu}] + A_\mu A_\nu[\frac{1}{4}\varphi^4 F^{\alpha\beta}F_{\alpha\beta} - \varphi(D_\alpha\partial^\alpha\varphi)], \\
\hat{R}_{5\mu} &= -\frac{1}{2}\varphi^2(D_\lambda F^\lambda{}_\mu) - \frac{3}{2}\varphi\partial^\lambda\varphi F_{\lambda\mu} + A_\mu[\frac{1}{4}\varphi^4 F^{\alpha\beta}F_{\alpha\beta} - \varphi(D_\alpha\partial^\alpha\varphi)], \\
\hat{R}_{55} &= \frac{1}{4}\varphi^4 F^{\alpha\beta}F_{\alpha\beta} - \varphi(D_\alpha\partial^\alpha\varphi)
\end{aligned} \tag{3.3.8}$$

Now it can be seen that (3.3.8) and (3.3.1) reduce to

$$\begin{aligned}
\hat{R}_{\mu\nu} &= 0 \\
R_{\mu\nu} &= \frac{1}{2}\varphi^2 F_{\mu\lambda}F_\nu{}^\lambda + \frac{1}{\varphi}(D_\nu\partial_\mu\varphi), \\
\hat{R}_{5\mu} &= 0 \\
D_\lambda F^\lambda{}_\mu &= -\frac{3}{\varphi}\partial^\lambda\varphi F_{\lambda\mu}, \\
\hat{R}_{55} &= 0 \\
D_\alpha\partial^\alpha\varphi &= \frac{1}{4}\varphi^3 F^{\alpha\beta}F_{\alpha\beta}.
\end{aligned} \tag{3.3.9}$$

In order to obtain the 4-D energy-stress tensor from 5-D Einstein tensor, we have to calculate

$$\hat{G}^{\mu\nu} = \hat{R}^{\mu\nu} - \frac{1}{2}\hat{g}^{\mu\nu}\hat{R}. \tag{3.3.10}$$

Starting with the Ricci scalar

$$\hat{R} = \hat{g}^{\mu\nu}\hat{R}_{\mu\nu} + \hat{g}^{\mu 5}\hat{R}_{\mu 5} + \hat{g}^{5\nu}\hat{R}_{5\nu} + \hat{g}^{55}\hat{R}_{55} \qquad (3.3.11)$$

this yields,

$$\hat{R} = R - \frac{1}{4}\varphi^2 F^{\alpha\beta}F_{\alpha\beta} - \frac{2}{\varphi}(D_\alpha \partial^\alpha \varphi), \qquad (3.3.12)$$

where R is the well known 4-D Ricci scalar.

The expressions for the contravariant components of $\hat{R}^{\mu\nu}$ and $\hat{G}^{\mu\nu}$ turn out to be more revealing:

$$\hat{G}^{\mu\nu} = \hat{R}^{\mu\nu} - \frac{1}{2}\hat{g}^{\mu\nu}\hat{R}. \qquad (3.3.13)$$

Using (3.3.4) and (3.3.6) one can obtain

$$\hat{R}^{\mu\nu} = \hat{g}^{\mu A}\hat{g}^{\nu B}\hat{R}_{AB},$$

$$\hat{R}^{\mu\nu} = R^{\mu\nu} - \frac{1}{2}\varphi^2 F^\mu{}_\lambda F^{\nu\lambda} - \frac{1}{\varphi}(D^\nu \partial^\mu \varphi), \qquad (3.3.14)$$

$$\hat{G}^{\mu\nu} = R^{\mu\nu} - \frac{1}{2}\varphi^2 F^\mu{}_\lambda F^{\nu\lambda} - \frac{1}{\varphi}(D^\nu \partial^\mu \varphi)$$
$$-\frac{1}{2}g^{\mu\nu}(R - \frac{1}{4}\varphi^2 F^{\alpha\beta}F_{\alpha\beta} - \frac{2}{\varphi}(D_\alpha \partial^\alpha \varphi)) \qquad (3.3.15)$$

Rearranging this, yields,

$$\hat{G}^{\mu\nu} = [R^{\mu\nu} - \frac{1}{2}g^{\mu\nu}R] - \frac{1}{2}\varphi^2[F^{\mu}{}_{\lambda}F^{\nu\lambda} - \frac{1}{4}g^{\mu\nu}F^{\alpha\beta}F_{\alpha\beta}]$$
$$-\frac{1}{\varphi}[(D^{\nu}\partial^{\mu}\varphi) - g^{\mu\nu}(D_{\alpha}\partial^{\alpha}\varphi)]$$
(3.3.16)

Now we can write Einstein's equations with the stress tensor

$$G^{\mu\nu} = R^{\mu\nu} - \frac{1}{2}g^{\mu\nu}R = T^{\mu\nu}{}_{EM} + T^{\mu\nu}{}_{S},$$
(3.3.17)

where

$$T^{\mu\nu}{}_{EM} = \frac{1}{2}\varphi^2[F^{\mu}{}_{\lambda}F^{\nu\lambda} - \frac{1}{4}g^{\mu\nu}F^{\alpha\beta}F_{\alpha\beta}],$$
(3.3.18a)

and

$$T^{\mu\nu}{}_{S} = \frac{1}{\varphi}[(D^{\nu}\partial^{\mu}\varphi) - g^{\mu\nu}(D_{\alpha}\partial^{\alpha}\varphi)]$$
(3.3.18b)

The first equation is the stress tensor for the electromagnetic field and the second is the stress tensor for the scalar field $\varphi$.

Choosing the scalar field $\varphi$ to be constant in (3.3.17), the second equation of (3.3.9) gives us Einstein's equations with the electromagnetic stress tensor and Maxwell's equations respectively:

$$G^{\mu\nu} = T^{\mu\nu}{}_{EM}$$
(3.3.19)

and

$$D_{\lambda}F^{\lambda}{}_{\mu} = 0.$$
(3.3.20)

Also using (3.3.9) and (3.3.18) we can conclude that $|\vec{E}| = |\vec{B}|$ because

$$F^{\alpha\beta}F_{\alpha\beta} = 0.\tag{3.3.21}$$

But we didn't consider the scalar field yet. After Kaluza, Thiry and Jordon [15],[16] mentioned about the scalar field. Turning back to the third equation of (3.3.7) we define the d'Alembertian operator as

$$\Box = D_\alpha D^\alpha,\tag{3.3.22}$$

and is equivalent to the covariant divergence of $\partial^\alpha \varphi$ while $\varphi$ being a scalar field. So we can write

$$\Box \varphi = D_\alpha D^\alpha \varphi \equiv D_\alpha \partial^\alpha \varphi.\tag{3.3.23}$$

From (3.3.9)

$$\Box \varphi = 0,\tag{3.3.24}$$

which is just the Klein-Gordon equation for a scalar field. Thus the scalar field $\varphi$ satisfies the Klein-Gordon equation. This property of $\varphi$ made the Kaluza-Klein theory well quantized at that time. But after the development of the quantum field theory Klein-Gordon equation lost its popularity.

## HORIZONTAL LIFT BASE

We shall introduce an anholonomic basis called the "horizontal lift basis" (HLB), not only for the convenience of subsequent calculations, but also due to the fact that the structure of the theory is well displayed in this basis.

HLB is defined in the following way:

$$\hat{\theta}^\mu = dx^\mu \tag{3.4.1}$$

$$\hat{\theta}^5 = dy + \kappa A_\mu(x) dx^\mu \tag{3.4.2}$$

$$\hat{g}_{\hat{\mu}\hat{\upsilon}} = \begin{pmatrix} g_{\mu\upsilon} & 0 \\ 0 & -1 \end{pmatrix}. \tag{3.4.3}$$

Now $\hat{e}_{\hat{\mu}}$ dual to $\hat{\theta}^\mu$ are defined as

$$e_\mu = (\partial/\partial x^\mu) - \kappa A_\mu (\partial/\partial y), \tag{3.4.4}$$

$$e_5 = (\partial/\partial y). \tag{3.4.5}$$

Because HLB is anholonomic it has non-vanishing commutators:

$$[e_\mu, e_\upsilon] = -\kappa F_{\mu\upsilon} (\partial/\partial y) \tag{3.4.6}$$

$$[\hat{e}_\mu, \hat{e}_\upsilon] = 0 \tag{3.4.7}$$

By this way non-vanishing connections are found to be

$$\Gamma_{\mu\upsilon\lambda}, \quad \hat{\Gamma}_{\mu\upsilon 5} = \hat{\Gamma}_{\mu 5\upsilon}, \quad \Gamma_{5\mu\upsilon}. \tag{3.4.8}$$

In the next chapter we are going to use HLB in order to make the calculations easier and more revealing.

# CHAPTER 4

# KILLING-YANO SYMMETRY IN

# 5-D KALUZA-KLEIN THEORY

## 5-D Metric with HLB

Taking 5D-Kaluza-Klein structure into account we can extend any metric into 5 dimensions.

$$\hat{g}_{\hat{\mu}\hat{\upsilon}} = \begin{pmatrix} g_{\mu\upsilon} - \kappa^2 A_\mu A_\upsilon & -\kappa^2 A_\mu \\ -\kappa^2 A_\upsilon & -1 \end{pmatrix} \qquad (4.1.1)$$

Using HLB, a simpler form can be obtained

$$\hat{g}_{\hat{\mu}\hat{\upsilon}} = \begin{pmatrix} g_{\mu\upsilon} & 0 \\ 0 & -1 \end{pmatrix} \qquad (4.1.2)$$

## KILLING-YANO EQUATION FOR 5-D KK METRIC

Since the well known Killing-Yano equation for valence 2 tensors has the form

$$D_A \hat{f}_{BC} + D_B \hat{f}_{AC} = 0. \tag{4.2.0}$$

Therefore, the crucial objects that contribute to the Killing-Yano equations are the connections coming from covariant derivatives via the structure of the metric.

Expansion of the equations to 5-dimensions yields eight classes of equations:

$$D_\mu \hat{f}_{\nu\lambda} + D_\nu \hat{f}_{\mu\lambda} = 0 \tag{4.2.1}$$

$$D_\mu \hat{f}_{\nu 5} + D_\nu \hat{f}_{\mu 5} = 0 \tag{4.2.2}$$

$$D_\mu \hat{f}_{5\lambda} + D_5 \hat{f}_{\mu\lambda} = 0 \tag{4.2.3}$$

$$D_\mu \hat{f}_{55} + D_5 \hat{f}_{\mu 5} = 0 \tag{4.2.4}$$

$$D_5 \hat{f}_{\nu\lambda} + D_\nu \hat{f}_{5\lambda} = 0 \tag{4.2.5}$$

$$D_5 \hat{f}_{\nu 5} + D_\nu \hat{f}_{55} = 0 \tag{4.2.6}$$

$$D_5 \hat{f}_{5\lambda} + D_5 \hat{f}_{5\lambda} = 0 \tag{4.2.7}$$

$$D_5 \hat{f}_{55} + D_5 \hat{f}_{55} = 0 \tag{4.2.8}$$

It's easy to see that (4.2.4) and (4.2.6) are the same and (4.2.8) is identically zero.

Considering the anti-symmetry of $\hat{f}_{\mu\lambda}$'s, it can be seen that there are 10 independent components of the KY tensor in 5-D.

Expanding the equations and putting the Kaluza-Klein connections with fifth components into the equations and using the cylinder condition we get the following equations:

$$\partial_\mu \hat{f}_{\nu\lambda} + \partial_\nu \hat{f}_{\mu\lambda} - 2\hat{\Gamma}^\delta_{\mu\nu}\hat{f}_{\delta\lambda} - \hat{\Gamma}^\delta_{\lambda\mu}\hat{f}_{\nu\delta} - \hat{\Gamma}^\delta_{\lambda\nu}\hat{f}_{\mu\delta} - \varphi F_{\mu\nu}\hat{f}_{5\lambda} + \frac{1}{2}\varphi F_{\lambda\mu}\hat{f}_{5\nu} + \frac{1}{2}\varphi F_{\lambda\nu}\hat{f}_{5\mu} = 0 \qquad (4.2.9)$$

$$\partial_\mu \hat{f}_{5\nu} + \partial_\nu \hat{f}_{5\mu} - 2\hat{\Gamma}^\delta_{\mu\nu}\hat{f}_{5\delta} - \frac{1}{2}\varphi F^\delta{}_\mu \hat{f}_{\nu\delta} - \frac{1}{2}\varphi F^\delta{}_\nu \hat{f}_{\mu\delta} = 0 \qquad (4.2.10)$$

$$\partial_\mu \hat{f}_{5\nu} + \frac{1}{\varphi}\partial_\nu \varphi \hat{f}_{5\mu} - \frac{1}{\varphi}\partial_\mu \varphi \hat{f}_{5\nu} + \varphi F^\delta{}_\mu \hat{f}_{\delta\nu} - \hat{\Gamma}^\delta_{\mu\nu}\hat{f}_{5\delta} + \frac{1}{2}\varphi F^\delta{}_\nu \hat{f}_{\mu\delta} = 0 \qquad (4.2.11)$$

$$\frac{1}{2}\varphi F^\delta{}_\mu \hat{f}_{\delta 5} + \frac{1}{\varphi}\partial^\delta \varphi \hat{f}_{\mu\delta} = 0 \qquad (4.2.12)$$

$$\partial_\mu \hat{f}_{5\nu} - \frac{1}{\varphi}\partial_\mu \varphi \hat{f}_{5\nu} + \frac{1}{\varphi}\partial_\nu \varphi \hat{f}_{5\mu} + \varphi F^\delta{}_\mu \hat{f}_{\delta\nu} + \frac{1}{2}\varphi F^\delta{}_\nu \hat{f}_{\mu\delta} - \hat{\Gamma}^\delta_{\mu\nu}\hat{f}_{5\delta} = 0 \qquad (4.2.13)$$

$$\frac{1}{\varphi}\partial^\delta \varphi \hat{f}_{\delta\lambda} + \frac{1}{2}\varphi F^\delta{}_\lambda \hat{f}_{5\delta} = 0 \qquad (4.2.14)$$

It can easily be seen that equations (4.2.12) and (4.2.14) are the same.

So we have five classes of independent equations.

For the 5-D case solution of Killing–Yano tensors can only be solved simultaneously with appropriate Maxwell stress tensors.

### REDUCED KILLING –YANO EQUATIONS FOR THE MINKOWSKI METRIC

After obtaining the KY equations for a general curved metric with, one can foresee that those equations become much simpler if we omit gravity. Using a Minkowskian metric yields 4-D connections to be zero. Then we have the following equations:

$$\partial_\mu \hat{f}_{\nu\lambda} + \partial_\nu \hat{f}_{\mu\lambda} - \varphi F_{\mu\nu} \hat{f}_{5\lambda} + \frac{1}{2}\varphi F_{\lambda\mu} \hat{f}_{5\nu} + \frac{1}{2}\varphi F_{\lambda\nu} \hat{f}_{5\mu} = 0 \tag{4.2.15}$$

$$\partial_\mu \hat{f}_{5\nu} + \partial_\nu \hat{f}_{5\mu} - \frac{1}{2}\varphi F^\delta{}_\mu \hat{f}_{\nu\delta} - \frac{1}{2}\varphi F^\delta{}_\nu \hat{f}_{\mu\delta} = 0 \tag{4.2.16}$$

$$\partial_\mu \hat{f}_{5\nu} + \frac{1}{\varphi}\partial_\nu \varphi \hat{f}_{5\mu} - \frac{1}{\varphi}\partial_\mu \varphi \hat{f}_{5\nu} + \varphi F^\delta{}_\mu \hat{f}_{\delta\nu} + \frac{1}{2}\varphi F^\delta{}_\nu \hat{f}_{\mu\delta} = 0 \tag{4.2.17}$$

$$\frac{1}{2}\varphi F^\delta{}_\mu \hat{f}_{\delta 5} + \frac{1}{\varphi}\partial^\delta \varphi \hat{f}_{\mu\delta} = 0 \tag{4.2.18}$$

$$\partial_\mu \hat{f}_{5\nu} - \frac{1}{\varphi}\partial_\mu \varphi \hat{f}_{5\nu} + \frac{1}{\varphi}\partial_\nu \varphi \hat{f}_{5\mu} + \varphi F^\delta{}_\mu \hat{f}_{\delta\nu} + \frac{1}{2}\varphi F^\delta{}_\nu \hat{f}_{\mu\delta} = 0. \tag{4.2.19}$$

Letting all the KY tensor components except for the fifth components to be constants makes us focus on the equation (4.2.15). It becomes

$$-\varphi F_{\mu\nu} \hat{f}_{5\lambda} + \frac{1}{2}\varphi F_{\lambda\mu} \hat{f}_{5\nu} + \frac{1}{2}\varphi F_{\lambda\nu} \hat{f}_{5\mu} = 0. \tag{4.2.20}$$

Contracting this by $\eta^{\mu\nu}$ (4.2.20) reads as

$$\varphi F_\lambda{}^\delta \hat{f}_{5\delta} = 0. \tag{4.2.21}$$

Using (4.2.21) in (4.2.18), we have

$$\frac{1}{\varphi}\partial^\delta \varphi \hat{f}_{\mu\delta} = 0. \tag{4.2.22}$$

Equation (4.2.22) implies that there are two cases:

i) $\quad \partial^\delta \varphi = 0$.

which means $\varphi$ is a constant and

ii) $\quad \det \hat{f}_{\mu\delta} = 0$.

These are the conditions for $\varphi$ and $\hat{f}_{\mu\delta}$. The condition for $F_{\mu\nu}$, the electromagnetic tensor, comes from (4.2.21) itself. If $\hat{f}_{5\delta}$ is not zero, then

$$\det F_{\mu\nu} = 0.$$

To summarize, if $\hat{f}_{5\delta}$'s are to survive and $\varphi$ not being constant, then we have two necessary conditions. These are:

1. $\hat{f}_{\mu\nu}$'s are constant.
2. $\det F_{\mu\nu} = \det \hat{f}_{\mu\nu} = 0$.

For simplicity we can make more assumptions, such as taking all KY tensor components to be constants, but still the above conditions will prevail.

# CHAPTER 5

# CONCLUSION

The first part of this thesis is dedicated to the understanding of Killing-Yano tensors. Their relations with the Killing tensors of valence two and three are presented. It is also pointed out that the recent revival of interest for KY tensors mostly comes from the fact that they are shown to be intimately related to supercharges for spinning point particles on curved backgrounds. Although Dietz and Rudiger [25] have given a classification of the line elements admitting KY tensors of valence two and four, each such manifold should be investigated independently, in order to find the explicit forms of the metric functions and the explicit expressions of the KY tensors. To exemplify, we have reviewed the case for the pp-wave metric, and we have found solutions for KY tensors of valence two and three. Our strategy was to find the maximum number of non-zero KY tensor components with minimum restrictions on the metric function of the pp-wave metric. From those valence two and three KY tensors we have constructed the corresponding second rank Killing tensors.

In the second part of this thesis we have reviewed the 5-D Kaluza-Klein theory. The horizontal lift base is introduced in order to make the 5-D Kaluza-Klein equations more revealing. Then we wrote KY equations in five dimensions with the background metric chosen to be that of the KK theory. In this way we have accommodated KY equations into the KK theory. After implementing the conventional KK reduction mechanism, the 5-D KY equations of (4.2.0) renders five sets of equations in 4D, totally giving 116 equations due to symmetries. At this point it is not apparent weather these equations admit physically meaningful solutions. To tackle this problem, after the KK reduction, we have chosen Minkowskian 4-D metric and considered the 4-D KY tensors as constant tensors and investigated the necessary conditions on the KY tensors and the EM field tensor. It is crucial that not all $\hat{f}_{5\mu}$ are zero; otherwise 5-D KY à la KK would be meaningless. Then we have found that if the scalar field $\varphi$ is not constant, the determinant of $\hat{f}_{\mu\nu}$ and the determinant the EM field tensor $F_{\mu\nu}$ must both vanish. If $\varphi$ is constant, then the determinant of $\hat{f}_{\mu\nu}$ must be non zero, but det $F_{\mu\nu}$ still has to vanish.

As to a future study there are various points to be investigated. First of all, we may release the condition on the $\hat{f}_{\mu\nu}$, and let them to be coordinate dependent, along with an arbitrary curved background metric satisfying the conditions of [25], and investigate the necessary conditions on the EM field tensor and the consistency conditions of the resultant equations.

Then it may be possible to look for explicit solutions for the KY tensors and the explicit forms of the metric functions as we did in Sec. (2.4).

# REFERENCES

[1]   K. Yano, Ann. Math., 55 (1952) 328

[2]   R.H. Rietdijk and J.W. van Holten, hep-th/9511166v1

[3]   B. Carter , Phys. Rev., 174 (1968) 1559

[4]   K. Rosquist, gr-qc/9410011

[5]   I. A. Barducci, R. Casalbuoni and L. Lusanna, *Nuovo Cimento*, A35 (1976) 377;
R.Casalbuoni, Nuovo Cimento, A33 (1976) 115; *id,* A33 (1976) 389

[6]   D. Baleanu, S. Baskal, gr-qc/0206045v1

[7]   W. Kundt, Z. Phys., 163, (1961) 77

[8]   H. Kuyrukcu, **Ph D Thesis**, M.E.T.U 2010

[9]   C. W. Misner, **Gravitation** 1971

[10]  J. D. Jackson, **Classical Electrodynamics** 3rd Ed. 1999

[11]  H. C. Lee, **An Introduction to Kaluza-Klein Theories** 1983

[12]  D. Baleanu, S. Baskal, Int. J. Mod. Phys. A 3737 (2002) 17

[13]  D. Ray, **Introducing Einstein's Relativity** 1992

[14]  T. Kaluza, Sitz. Press Akademie Wiss. Phys. Math. K1, (1921) 966

[15]  P. Jordon, Ann. Der Phys. (Leipzig), 1 (1947) 219

[16]  Y. Thiry, Acad. Sci. (Paris), 226 (1948)16

[17]  O. Klein, Zeits. Phys., 37 (1926) 895

[18]  O. Klein, Nature, 118 (1926) 516

[19]  G. Nordström, Phys. Zeitschr., 15 (1914) 504

[20]  J. M. Overduin, P. S. Wesson, gr-gc/9805018

[21]  R. Geroch, J. Math. Phys., 13 (1972) 394

[22]  P. Sommers, J. Math. Phys., 14 (1973) 787

[23]  D. Kramer, H. Stephani, E. Herlt, M. MacCallum, C. Hoenselaers, E. Herlt
**Exact Solutions of Einstein's Field Equations** 2nd Ed. 2003

[24]  D. Garfinkle, E. N. Glass, arXiv:1003.0019v2 [gr-qc]

[25]  W. Dietz, R. Rudiger, Proc. R. Soc. Lond.,  A 375 (1981) 361

[26]  O. Acik, U. Ertem, M. Onder, A. Vercin, Gen. Relativ. Gravit., 42 (2010) 2543

[27]   O. Acik, U. Ertem, M. Onder, A. Vercin, Class. Quantum Grav., 26 (2009) 075001